\documentstyle[epsf,graphicx]{mn}

\def\etal{{et al.\ }}

\def\x2{$\chi^{2}$}

\def\asca{{\it ASCA }}
\def\rosat{{\it ROSAT }}

\def\x2{$\chi^{2}$}
\def\lunits{$\rm{erg\,s^{-1}}$}
\def\funits{$\rm{erg\,s^{-1}\,cm^{-2}}$}
\def\cunits{$\rm{cm^{-2}}$}

\newbox\grsign \setbox\grsign=\hbox{$>$} \newdimen\grdimen \grdimen=\ht\grsign
\newbox\simlessbox \newbox\simgreatbox \newbox\simpropbox
\setbox\simgreatbox=\hbox{\raise.5ex\hbox{$>$}\llap
     {\lower.5ex\hbox{$\sim$}}}\ht1=\grdimen\dp1=0pt
\setbox\simlessbox=\hbox{\raise.5ex\hbox{$<$}\llap
     {\lower.5ex\hbox{$\sim$}}}\ht2=\grdimen\dp2=0pt
\setbox\simpropbox=\hbox{\raise.5ex\hbox{$\propto$}\llap
     {\lower.5ex\hbox{$\sim$}}}\ht2=\grdimen\dp2=0pt

\begin{document}

\title[ X-ray observations of Superantannae] {X--ray observations of the
Ultraluminous infrared galaxy IRAS19254-7245 (The Superantennae)}

\author[A. Pappa, I. Georgantopoulos  and G.C. Stewart]
       {A. Pappa$^{1}$, I. Georgantopoulos$^{2}$ and G.C. Stewart$^{1}$ \\
 Department of Physics and Astronomy, University of Leicester, 
Leicester, LE1 7RH \\
 Astronomical Institute, National Observatory of Athens, Lofos Koufou, 
 Palaia Penteli, 
15236, Athens, Greece } 

\maketitle

\label{firstpage}

\begin{abstract}
We present ROSAT HRI and ASCA observations of the 
well known ULIRG IRAS19254-7245 (the Superantennae).
The object is not detected 
by ROSAT yielding a 3$\sigma$ upper limit of
$L_x \sim 8\times10^{41}$\lunits in the 0.1-2 keV band. 
 However, we obtain a  clear detection by ASCA
 yielding a luminosity in the 2-10 keV band of 
 $2 \times 10^{42}$ \lunits. Its X-ray spectrum  
is very hard, equivalent to a photon index  of 
 $\Gamma=1.0\pm0.35$.  
 We therefore, attempt to model the 
 X-ray data with a "scatterer" model in which the intrinsic  
  X-ray emission along our line of sight  is 
 obscured by an absorbing screen 
 while some fraction, f, is scattered 
 into our line of sight by an ionized medium; 
 this is the standard model for the 
 X-ray emission in obscured (but non Compton-thick) 
 Seyfert galaxies.  
 We obtain an absorbing column of 
 $2\times 10^{23}$ $\rm cm^{-2}$ for a power-law 
 photon index of $\Gamma=1.9$, an order 
 of magnitude  above the column estimated on the basis 
 of optical observations; the percentage of the 
 scattered emission is high ($\sim$ 20 per cent).
 Alternatively, a model where most of the X-ray emission 
 comes from reflection on a Compton thick torus 
 ($N_H>10^{24} \rm cm^{-2}$) cannot be ruled out. 
 We do not detect an Fe line at 6.4 keV; however, 
  the upper limit (90$\%$) 
 to the equivalent width of 
 the 6.4 keV line is high  ($\sim$3 keV). 
 All the above  suggest 
 that most of the X-ray emission originates in an
 highly obscured Seyfert-2 nucleus.

\end{abstract}


\section{INTRODUCTION}

The IRAS Ultraluminous Infrared Galaxies (ULIRGs), $L_{IR}\geq10^{12}L_\odot$,
 are among the most luminous objects in the Universe. ULIRGs  emit
most  of their energy in the far-infrared
 (see Sanders \& Mirabel 1996 for a recent review).
 The nature of this powerful emission has been 
 hotly debated. The far-IR emission is clearly 
 produced by thermal reradiation by dust. 
 However, it remains unclear whether the  heating of the dust is 
 due to  a hidden AGN and/or massive star-forming regions. 
 Optical and near-IR imaging surveys (eg Duc, Mirabel \& Maza 1997) 
 show that most ULIRGs are close interacting or merging systems.
 Optical spectroscopic surveys show that a large fraction (about 30 per cent) 
 of ULIRGs are associated with Seyfert nuclei (Kim 1995, Duc et al. 1997, 
 Sanders et al. 1998). The rest appear to host either 
 starforming or LINER nuclei. 
 Lutz et al. (1999) have reached similar conclusions 
 using far-IR spectroscopy from the ISO mission.  
 The ULIRGs which have AGN like 
 spectra, may be associated with the long sought   
 population of type-2 QSOs, ie high redshift 
 AGN with narrow-line Seyfert-2 type spectra but 
 bolometric luminosities comparable with those of QSOs. 
X--ray observations at high energies which remain largely 
 unaffected by absorption are vital  in 
checking for the presence of an AGN in the nuclei of the ULIRG.
 
Recently, a handful of ULIRGs have been observed with the ASCA 
X-ray satellite: IRAS09104+4109 (Fabian \etal 1994), 
IRAS15307+3252, IRAS 20460+1925 (Ogasaka et al. 1997), 
IRAS 23060+0505 (Brandt et al. 1997), NGC6240, Arp220, Mrk 273, 
and Mrk231 (Iwasawa \& Comastri 1998). All have been  
detected apart from IRAS15307+3252. Most  of the above 
ULIRGs have  2-10 keV luminosities $L_x>10^{42}$ 
 \lunits clearly suggesting the presence of 
 a buried AGN. These high X-ray luminosity objects 
  have power-law spectra absorbed by large neutral  hydrogen columns, 
  typically $N_H>10^{22}$ $\rm cm^{-2}$; their spectra also 
 show evidence for an Fe-K line at 6.4 keV which again 
 suggests the presence of large amounts of neutral matter 
 near the nucleus. However, the origin of the X-ray emission 
 in Arp 220 and Mrk231 appears to be thermal (Iwasawa \& Comastri 1998). 
    
\subsection {IRAS19254-7245 (The Superantannae)}

The Superantennae ($L_{IR}$=$1.1\times 10^{12} L_\odot$)  is a remarkable
ULIRG with a redshift of z=0.0617 (Mirabel, Lutz \& Maza 1991).
It presents giant tails extending to an unprecedented size of 350 kpc. 
These emanate from a merger of two giant gas-rich galaxies whose  
 nuclei are separated by 10 kpc. 
 The southern nucleus is  heavily obscured ($A_V\sim4-5$)
and the optical observations give  a Seyfert 2 classification
 (Mirabel et al. 1991). Mid-infrared spectroscopy   (Lutz, Veilleux
 \& Genzel 1999) of the 
 superantennae again suggest  an AGN classification.  
Despite the  classification of the southern galaxy 
 as a Seyfert-2 on the basis of optical and far-IR spectra,
  spectropolarimetric observations of the 
 Superantennae (Heisler \etal 1997), 
 revealed no scattered broad line component. The authors
attributed this  to geometric effects. According to their
model the scattering particles, which produce the observed polarised
lines, must lie very close to or within the plane of the torus (see
also Miller $\&$ Goodrich 1990) so that if we see the galaxy edge-on,
we can't see the polarised flux. 
At near-IR wavelengths the Seyfert nucleus dominates the emission; at
10$\mu$m it is more than 5 times brighter than the northern component
(Sanders $\&$ Mirabel 1996) and it is likely to be the source of the
far-infrared emission detected by IRAS (Mirabel \etal 1991).
Melnick and Mirabel (1990)  reported the presence of a huge 
amount of molecular gas $M_{CO}=3\times10^{10}M_\odot$, 
10 times larger than  in our Galaxy. The mass of the
ionised gas in the emission-line regions is $M_g=1.1\times10^8M_\odot$
(Colina \etal 1991). The latter authors find that the starburst
produces too few high energy photons to explain the observed line
intensities and thus infer that the Superantennae must contain
a luminous $> 10^{45}$ \lunits AGN. 
Rush \etal (1996) presented results from the ROSAT
All-Sky Survey observation of the Superantannae. They did not detect the
object and the $2\sigma$ upper limit to the count rate is $<0.03$
counts/sec. The
estimated upper limit to the 0.1-2 keV luminosity is $4.5\times10^{42}$\,\lunits
(using a power-law of $\Gamma=2.3$ and assuming Galactic absorption).
Here,  we present the analysis of ROSAT HRI and ASCA data of the
Superantennae galaxy. The latter are the  first observations in the hard
X--ray band and as such  are important in testing  models
featuring a highly obscured, high luminosity nucleus.

\section{OBSERVATIONS AND DATA REDUCTION}

The Superantennae were observed with ASCA on 16/10/96. The net
exposure time for each GIS is $\sim30$ ksec, while the SIS0 net
exposure time is $\sim22$ ksec and the SIS1  is
$\sim10$ ksec.
We have used the ``Revision 2'' processed data from the Goddard Space
Flight Center (GSFC) and data reduction was performed using FTOOLS.
A circular extraction cell for the source of 3 arcminute 
radius has been used.
Background counts were estimated from source-free annuli centered on
the source cell.   
The observed flux in the 2-10 keV band is
$f_{2-10 \rm keV}\simeq2.4\times10^{-13}$
\funits \, while the one in the 1-2 keV band is 
$f_{1-2\rm keV}=3.2\times10^{-14}$\funits, 
  assuming  the best-fit power-law model of section  3 
 ($\Gamma=1$).

ROSAT observed the Superantennae with HRI for $\sim8$ ksec
  between 17/04/1993 and 20/04/1993. 
 There were no X--rays detected and the $3\sigma$
upper limit to emission from a point source
obtained in the 0.1-2 keV band is $3.2 \times 10^{-3}$ counts/sec, 
which correspond to a flux of 
$f_{\rm 1-2 keV}=9.5\times10^{-14}$\funits,  
 well above the flux obtained by \asca.    
 
Throughout this paper we adopt $ H_\circ=70$ $\rm km s^{-1} Mpc^{-1}$
and $q_o=0.5$.

\section{ SPECTRAL ANALYSIS}

The spectral analysis was carried out using XSPEC v10.
 We bin the data so that there are at least 20 counts 
 in each bin (source plus background). 
Quoted errors to the best-fitting spectral parameters are 90 per cent
confidence regions for one parameter of interest.
We first performed spectral fitting, allowing the normalisation for
each ASCA detector to vary and we obtained reasonably consistent
results. We have therefore jointly fitted the spectra from all four
detectors, tying their normalisations together.
The results of the spectral fits are given in table 1. 
Entries with no associated uncertainties were fixed at this 
 value during the spectral fit. 

\subsection{Obscured AGN Models}
Given that the optical and mid-infrared observations of the
Superantannae suggest an AGN classification we fit the GIS and SIS
data with a simple power law model with absorption. 
We obtain a flat slope of 1.0$\pm0.35$ with no requirement for absorption 
above the Galactic. The best-fit model together with the data points 
 and the data to model ratio are plotted in Fig. 1
Data have been rebinned for clarity. 

This slope is much flatter than is seen in radio quiet AGN and the low
absorption is not consistent with the $A_{\rm V}$ estimates from the
narrow lines or the absence of broad line emission.   
A hard continuum is, however, often seen 
in  Seyfert 2 galaxies (e.g. Georgantopoulos  \etal 1999 for
 Markarian 3). While the simple power-law model is formally acceptable
we have therefore also tried  a ``scatterer'' model in which 
the X-ray source  is covered by an absorption screen and a
fraction of the 
X-ray emission is
scattered into our line of sight. Fixing the slope at
the 1.9 value, which is typical of the underlying continuum of Seyfert 
galaxies, we
obtain an absorbing column of $2\times10^{23}$\cunits \,and
$\chi^2=36.8$ for 34 degrees of freedom (dof), a marginal improvement on the
simple power-law model. 
Using this model we derive an intrinsic 2-10 keV
luminosity of  $4.3\times10^{42}$\lunits. 

Most Seyfert galaxies show strong narrow iron K emission lines and these 
features are usually enhanced in Seyfert 2s. 
When we add a Gaussian emission line to the scatterer  model, 
constraining  its
energy  and width at 6.4 (rest-frame) and 0.01 keV, 
we do not obtain a significant detection of line emission
but can only set a 90\% upper limit to the equivalent width of
such a feature of $\sim3\rm keV$.

Alternatively, the flat power-law slope might suggest  that Compton reflection 
dominate in the ASCA energy range (eg Matt et al. 1996). 
 For that reason we have also tried a pure 
reflection model. This assumes that the reflection occurs from a slab
of neutral material subtending a solid angle of $2\pi$ sr to an X--ray
source located above the slab. Again fixing the intrinsic power-law
slope at 1.9 we obtain a completely unacceptable fit with  
 $\chi^2=75.3$ for 34 dof. Relaxing the constraint on the power-law slope 
does improve the statistical acceptability of this model
($\chi^2=39.9$ for 33 dof) but results in 
 power-law slope of $\Gamma=3.45^{+0.42}_{-0.27}$, which is much steeper
than is seen  in AGN other than some narrow-line Seyfert 1s.

Furthermore introducing a reflection model with a scattered power law when the
absorption is fixed at the Galactic value and $\Gamma$ at 1.9
gives a similar quality fit to the scatterer model
with  $\chi^2=36.4$ for 33 dof. We tie the spectral
index of the two power-law components to have the same value, since
this is what is expected by the elastic scattering of the primary
emission by the warm plasma. The addition of a Gaussian line at 6.4 keV,
(rest-frame), does not improve the fit
($\chi^2=35.5$ for 34 dof). 

\subsection{Thermal Models}
   
While the optical emission line strengths are highly suggestive of
a luminous AGN at the core of the Superantennae,
 some fraction  of the X-ray emission must come from 
 the numerous star-forming regions.  
 Recent 
studies of starburst galaxies such as NGC 3690 (Zezas, Georgantopoulos
 \& Ward 1998) provide 
evidence for a hard thermal component in the spectrum 
 of luminous star-forming galaxies. 
 For completeness
therefore we have investigated a model in which the emission 
consists of  two thermal (Raymond-Smith) components. 
 We have fixed the temperature of the soft component 
 to 0.8 keV following Zezas et al. (1998) while we 
 assumed a Galactic absorption and a solar metallicity. 
This again provides
a reasonable statistical description of the data ($\chi^2=41.0/34$ dof), albeit with a poorly
constrained temperature for the hard component 
($>12$ keV). Fixing the temperatures of the two components to the
values
determined by Zezas et al but allowing the thermal components to be
absorbed improves the fit of this model ($\chi^2 = 37.9$) and gives a 
column densities of $\sim 7 \times 10^{21}\rm cm^{-2}$, similar to that 
inferred from the value of $A_v$, and $\sim 10^{23} \rm cm^{-2}$.

\subsection{Mixed Models}

Given the evidence of a strong starburst in the object and the case for
an AGN nucleus, it is not unnatural to investigate a model in which 
X-ray emission from both contribute to our ASCA spectrum. We adopt
as a baseline for this model a  power-law 
 (with $\Gamma = 1.9$) to represent the nuclear
emission and  a Raymond-Smith thermal component to 
represent  the starburst. We allow the power-law component to have additional
absorption over and above that of the thermal component. 
Again we fix the temperature of the thermal component
at 0.8 keV. This results in an acceptable fit with 
$\chi^2 =43.2 $ for 34 dof with a column of $N_H=1.80\times
10^{22}$  $\rm cm^{-2}$  for the
power-law.  Again allowing the thermal component to be absorbed
improves the
fit ($\chi^2 = 36.7$ for 33 dof) with a column of $\sim 8 \times
10^{21}\rm cm^{-2}$. 

\begin{figure*}
\rotatebox{270}{\includegraphics[height=11.5cm]{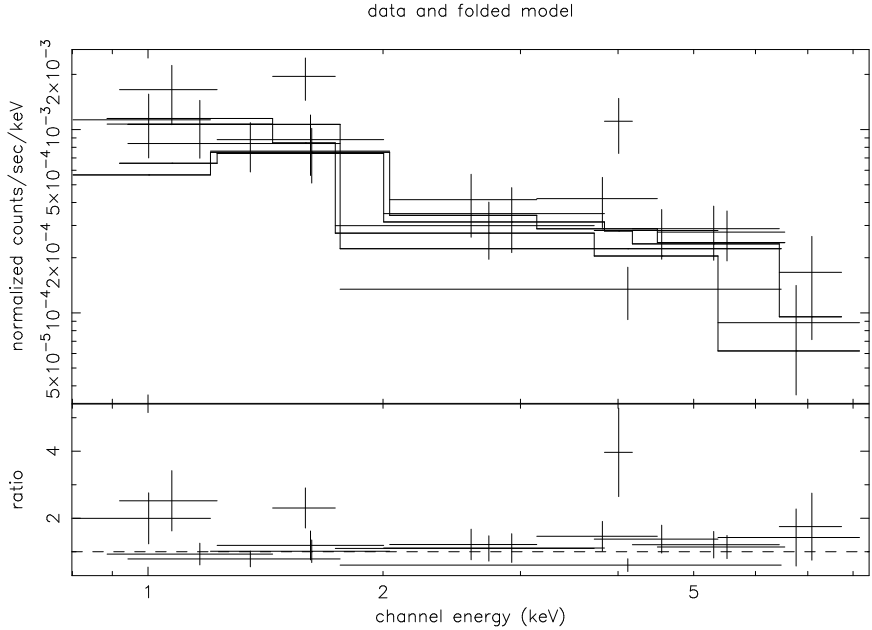}}
\caption{The best fit power-law of the ASCA X-ray spectrum of Superantennae. 
The top panel
shows the data with the model and the bottom panel shows the
data/model ratio.}
\end{figure*}

\section{DISCUSSION}

It becomes  evident that the 
 X-ray data alone are not sufficient to define the 
 nature of this enigmatic object. 
The X-ray data are well fit by both power-law 
 models and thermal models. 
 The temperatures in the latter are 
  not inconsistent with those found 
 by Zezas et al. (1998) in the case of the luminous IRAS galaxy NGC3690 
 which clearly shows no sign of AGN activity. 
 The derived luminosity ($L_x\sim 10^{42}$ \lunits)
 for the superantennae  
  is again comparable with that of NGC3690. 
 
 However, as  the optical spectrum reveals a  Seyfert-2 nucleus 
 in the 
 southernmost of the two merging nuclei, 
 (Colina et al. 1991), it is most probable that  
 a large fraction of the X-ray emission comes from 
 the active nucleus. 
 Indeed, the  hard  X-ray spectrum observed 
 ($\Gamma=1.0\pm 0.35)$ 
is reminiscent of highly absorbed Seyfert 2 galaxies ($N_H\sim 10^{23-24}$)
 $\rm cm^{-2}$  such as Markarian 3 (Georgantopoulos et al. 1999). 
 Alternatively, the possibility remains 
 that even in the \asca band we are not sensitive 
 to the intrinsic nuclear emission of the Superantennae. 
 This would require the source to 
 be totally obscured by a Compton thick absorber  
 ($N_H>10^{24}$ $\rm cm^{-2}$)
 in a similar fashion to NGC1068 (Ueno et al. 1994)
 or the Circinus (Matt et al. 1996).  Indeed, 
 both the Compton thin (power-law and scatterer model) and Compton thick   
 models (reflection plus a scattered power-law component) 
 give comparable $\chi^2$ values making it difficult to 
 distinguish between the two possibilities. 
In the  Compton-thin case, 
 the hard X-ray emission should 
 emerge from  an optically thin 
 screen (possibly in the form of the torus);
 the derived $N_H\sim 10^{23} \rm cm^{-2}$  is 
 comparable to those encountered in obscured 
 Seyfert galaxies  (see eg. Smith \& Done 1996, Risaliti \etal 1999). 
The absorption inferred by the optical extinction is
$9.3\times10^{21}$\cunits,  Colina \etal (1991); thus it appears that  
the X--ray absorbing column is at least
one order of magnitude higher than the optical one. 
 This discrepancy could then be explained either by assuming
that there is a strong absorption inside the broad line region  or that 
the IRAS19254-7245 gas-to-dust ratio is higher than  our Galaxy's.
In the Compton thin case  we find that the 
 power-law slope is compatible with the ``canonical'', intrinsic 
 AGN spectral index, $\Gamma=1.9$, (Nandra \& Pounds 1994).
 The observed soft X-ray emission is then 
 scattered radiation from a warm ionized medium
 (the temperature of this medium should be comparable with the 
 energy of the soft X-ray photons, so that elastic scattering 
 takes place). The above model has become 
 the standard model for the X-ray emission in 
 obscured Seyfert nuclei (Mushotzky, Done \& Pounds 1993).  
 Then the scattered component and the 
 intrinsic hard X-ray emission together with 
 high amounts of obscuration or reflection 
 can easily mimic a flat spectrum
 (eg as in the case of IRAS 23060+0505, Brandt et al. 1997).  
 In the case of IRAS 19254-7245, 
 we find that the amount of the scattered emission 
 towards our line of sight is about 20 per cent.  
 This is higher than typically found for the obscured Seyfert (Seyfert 1.9-2.0) nuclei,
 where the scattered component is usually of the order of few 
 percent. Colina \etal (1991) suggested that massive star formation
should take place  in order to explain the properties of the Superantannae
 (see also Mirabel \etal 1991).  Under this scenario the excess in the
 soft X-ray emission  could be explained by  a star
 forming component.
   The inferred soft X-ray luminosity,
 in the 1-2 keV band $L_x\sim 10^{41}$ \lunits is 
  quite typical of that expected in massive star-forming 
 galaxies (eg Zezas et al.  1998).    
 If instead we assume that all the soft X-ray emission 
 is mainly due to scattering, we can derive 
 interesting constraints on the location of the 
 scattering medium. If we require the ionisation 
 parameter of the scattering medium to be 
 $log\xi=3$ and the electron density $n_e=10^5 \rm cm^{-3}$,
 comparable to the values encountered in Mrk 3 (Griffiths 
 et al. 1998), we can derive the distance R of the 
 scattering medium from the continuum source. 
 We obtain $R\sim0.1$ pc suggesting that the scattering medium 
 lies close to the nucleus.  

 In the Compton thick case, the 
 X-ray emission comes from 
  both a reflection  and 
a scattered power-law component. 
 The geometry is similar to the 
 one described above in the case of the Compton thin model, 
 with the  exception that now 
 the torus is optically thick ($N_H>10^{24}$ $\rm cm^{-2}$).  
 We note that in both the Compton thick and thin
 case  we should detect a strong 
 Fe line at 6.4 keV.  Especially, in the Compton thick 
 case, the  equivalent width of the line could reach few keV
 as in eg the case of NGC6240 (Iwasawa \& Comastri 1999). 
 Our data do not show strong evidence for 
an Fe K emission line. However, the obtained upper limit to the
equivalent width (3 keV) does not in this case 
 rule out the Compton thick possibility.
 Vignati et al. (1999) recently showed, using {\it BeppoSAX} data
that the ULIRG NGC6240 hosts a Compton thick ($N_H\sim 2\times
 10^{24} \rm cm^{-2}$)   Seyfert-2 type nucleus.  
NGC6240 has a LINER classification in the optical while it is 
 classified as an HII galaxy on the basis of 
 IR spectroscopy (Lutz et al. 1999). 
 The LINER classification most probably comes from     
 low ionisation gas in a superwind. 
It is interesting 
 that although the superantennae has similar X-ray properties 
 to NGC6240, they present strong AGN characteristics 
 in the optical. This difference may be related to 
 the distribution of obscuring material in the galaxy. 

Further clues on the nature of the hidden AGN can be given by studying
the isotropic properties of the galaxy. Here, we will consider as
isotropic emission the
infrared (IR), the hard X-ray emission 
(in the case of Compton thin absorption) 
and the $[OIII]\lambda5007$ line emission produced in the
narrow line region, and thus free of viewing angle effects.
The advantage of studying isotropic properties, is that they act as an
indicator of the strength of the nuclear source.
 Maiolino et al. (1998) have proposed  that   
 the measurement of the unabsorbed hard
X--ray flux (2-10 keV) against the $[OIII]\lambda5007$ flux, 
is indeed a powerful diagnostic. Of course, this tool
 may be less efficient  
 in the case of ULIRGs which have 
 large amounts of dust.   
 Moreover,  although the 
line is emitted on  Narrow Line Region (NLR) scales, 
 the host galaxy disk
might obscure part of the NLR and should be corrected for the
extinction deduced from the Balmer decrement (Maiolino $\&$ Rieke 1995).
To estimate the corrected flux we used the observed 
$[OIII]\lambda5007$ flux from Colina \etal (1991) corrected for the
optical reddening using the following relation (Bassani \etal 1998):
\begin{equation}
F_{[OIII]cor}=F_{[OIII]obs} \times
[(H_{\alpha}/H_{\beta})/(H_{\alpha}/H_{\beta})_0]^{2.94}
\end{equation}
Assuming an intrinsic Balmer decrement $(H_{\alpha}/H_{\beta})_0=3$,
the $L_{\rm x(2-10keV)}/L_{[OIII]}$ ratio of Superantennae is $\sim0.01$
favouring the Compton-thick interpretation for IRAS 19254-7245.
 Then on the basis of the mean 
  $L_{\rm x(2-10keV)}/L_{[OIII]}$ ratio, 
the {\it intrinsic} luminosity of the Superantennae should exceed
 $10^{44}$ \lunits  suggesting that obscuration in our line of sight
prevents us from seeing any of the nuclear X-ray emission directly. 
Furthermore following Mulchaey \etal (1994) we use the IR to hard X-ray
flux ratio as an indicator. 
The infrared flux is given by  (Mulchaey \etal 1994):
\begin{equation}
F(IR)=S_{25{\mu}m}\times(\nu_{25{\mu}m})+S_{60{\mu}m}\times(\nu_{60{\mu}m})
\end{equation}
where $S_{\lambda}$ is the flux density at wavelength $\lambda$.
They showed that the expected $log(f_{IR}/f_x)$ (the IR to 
 {\it unabsorbed} hard X-ray
flux ratio) is $\sim0.9$. We find that for the superantennae this ratio
is $\sim3$, clearly indicating a deficit in the hard X-ray emission
compared to that expected from a Compton thin Seyfert 2 galaxy.
This again suggests a high column density in our light of
sight, which absorbs the photons in the \asca band.
We should note here, that the $f_{IR}/f_x$  ratio in the ULIRGs may be
high due to excess IR emission from a starburst component. 
 However, ISO observations suggest
an AGN classification for the Superantennae (Lutz \etal 1999), 
indicating that the AGN 
contribution to the infrared flux dominates over  
 the star-forming one.

\section{CONCLUSIONS}

 We have presented \rosat HRI and \asca X-ray observations 
 of the ULIRG IRAS 19254-7245 (the superantennae). 
 This object is not detected by the \rosat HRI.
 However, we detect hard X-ray emission with \asca 
  showing a very flat spectrum in the 1-10 keV band,
  reminiscent of the 
 spectra of highly obscured AGN locally. 
 Therefore, the  X-ray data indirectly argue in favour of the 
 existence of a supermassive black hole in IRAS 19254-7245. 
 In particular, the X-ray data  can be  modeled with 
 a power-law spectrum with a photon index slope 
 of $\Gamma=1.9$ emerging through 
 a Compton thin torus ($N_H\sim 10^{23} \rm cm^{-2})$.
 A scattered component (of the order of 20 per cent 
 of the nuclear component at 1 keV)  or alternatively 
 a large star-forming component ($L_x\sim 10^{41}$ \lunits) 
 is also present. We do not detect any Fe line at 6.4 keV with the 
 90 per cent upper limit on the equivalent width being 3 keV. 
 The X-ray data are also well modeled by a 
 Compton thick, reflection dominated model with some fraction of the
 nuclear emission scattered into the line of sight. 
  However, the limited quality of the present data 
 does not allow us to distinguish between the above two models.  
 Future observations with higher effective area missions 
 such as {\it Chandra} and {\it XMM} will be able to  provide 
 much better constraints on the origin of the 
 the soft X-ray emission as well as the 
 geometry of the obscuring material  
 thus shedding further light on the 
 nature of ULIRGs.

\section{Acknowledgments}

We would like to thank the anonymous referee for many useful 
 comments and suggestions. 
This research has made use of data obtained through the High Energy 
Astrophysics Science Archive Research Center Online Service, provided
by the NASA/Goddard Space Flight Center and the LEDAS online service,
provided by the University of Leicester.

\end{document}